\documentstyle[epsf]{elsart}

\def\be{\begin{equation}}
\def\ee{\end{equation}}

\def\simm#1{\mathop{\vtop{\ialign{##\crcr
        $\hfil\displaystyle{#1}\hfil$\crcr\noalign{\kern0.5pt\nointerlineskip}
        $\sim$\crcr\noalign{\kern0.5pt}}}}\limits}

\def\Fl{FLOPS}
\def\Ndx{{\tt Ndimx}}
\def\Ndy{{\tt Ndimy}}
\def\Ndz{{\tt Ndimz}}

\begin{document}
\begin{frontmatter}

\begin{flushright}
\normalsize
UTCCP-P-62  \\
March 1999  \\
\end{flushright}

\title{
Performance of lattice QCD programs on CP-PACS
}

\author[TsukubaP]{S. Aoki\thanksref{aoki}},
\author[TsukubaP,CCP]{R. Burkhalter\thanksref{ruedi}},
\author[TsukubaP,CCP]{K. Kanaya\thanksref{kanaya}},
\author[TsukubaP,CCP]{T. Yoshi\'e\thanksref{yoshie}},
\author[CCP,TsukubaC]{T. Boku\thanksref{boku}},
\author[Tokyo]{H. Nakamura\thanksref{nakamura}}
\and
\author[TsukubaC]{Y. Yamashita\thanksref{yaman}}

\address[TsukubaP]{Institute of Physics, University of Tsukuba, 
Tsukuba, Ibaraki 305-8571, Japan}

\address[CCP]{Center for Computational Physics, University of Tsukuba, 
Tsukuba, Ibaraki 305-8577, Japan}

\address[TsukubaC]{Institute of Information Sciences and Electronics, 
University of Tsukuba, Tsukuba, Ibaraki 305-8573, Japan}

\address[Tokyo]{Research Center for Advanced Science and Technology,
         The University of Tokyo, Komaba, Tokyo 153-8904, Japan}

\thanks[aoki]{\tt saoki@het.ph.tsukuba.ac.jp}
\thanks[ruedi]{\tt burkhalt@rccp.tsukuba.ac.jp}
\thanks[kanaya]{\tt kanaya@rccp.tsukuba.ac.jp}
\thanks[yoshie]{\tt yoshie@het.ph.tsukuba.ac.jp}
\thanks[boku]{\tt taisuke@is.tsukuba.ac.jp}
\thanks[nakamura]{\tt nakamura@hal.rcast.u-tokyo.ac.jp}
\thanks[yaman]{\tt yaman@is.tsukuba.ac.jp}

\begin{abstract}
The CP-PACS is a massively parallel MIMD computer with the theoretical peak
speed of 614 GFLOPS which has been developed for computational physics 
applications at the University of Tsukuba, Japan.  
We report on the performance of the CP-PACS computer measured during 
recent production runs using our Quantum Chromodynamics code 
for the simulation of quarks and gluons in particle physics.
With the full 2048 processing nodes, our code 
shows a sustained speed of 237.5 G\Fl\ for the heat-bath update of gluon
variables, 264.6 G\Fl\ for the over-relaxation update, and 325.3 G\Fl\ for 
quark matrix inversion with an even-odd preconditioned minimal residual 
algorithm.
\end{abstract}

\end{frontmatter}

\section{Introduction}

Quarks and gluons are the building blocks of a large number of 
elementary particles, collectively called hadrons, that include 
well-known particles such as protons and neutrons.  
A remarkable property of quarks and gluons is confinement,
that is while 
there is solid evidence that they exist within hadrons, they have 
never been observed in isolation in experiments.  The theoretical
principle governing the physical dynamics of quarks and gluons 
is described by a gauge field theory called quantum chromodynamics (QCD).

QCD is a highly non-linear quantum mechanical system in which the basic 
quark and gluon field degrees of freedom are defined at each point of 
four-dimensional space-time.  While the fluctuations of the fields with 
short wave length are weakly coupled, the coupling becomes stronger 
for longer wave lengths.  These features render an analytical solution
of QCD an impossible arduous task. Instead 
progress over the past two decades came from numerical simulations 
using a formulation of QCD on a four-dimensional space-time lattice, known as 
lattice QCD\cite{Wilson,intro}.

Approximating continuous space-time with a sufficiently fine
lattice necessarily requires a large lattice size $L$, with the consequence 
that the number of degrees of freedom increases as $L^4$.
When we increase $L$ we usually reduce the light quark mass 
such that it becomes closer to the physical value; 
this requires additional computations due to the critical slowing down. 
Taking these two factors into account, 
the amount of computing actually needed\cite{Sharpe} is considered to 
increase, at least, as fast as $L^{(8-10)}$.  
The numerical simulation of lattice QCD therefore requires 
significant computing power.   
On the other hand, quark and gluon 
fields interact only locally in space-time in QCD. Thus lattice QCD 
simulations are ideally suited for parallelism in the space-time 
coordinates.  

Exploiting this feature, a number of dedicated parallel computers has been 
developed since the 1980's aiming to advance lattice QCD 
simulations\cite{review}.  
The CP-PACS parallel computer is one of the latest efforts in this 
direction\cite{cppacs,other}. 
It is worth emphasizing, however, that the parallelism inherent in lattice 
QCD is shared by a large number of physics problems in which space-time or 
space fields are the basic dynamical variable.  Thus the overall objective 
of the CP-PACS Project is broader,  encompassing 
astrophysics and condensed matter applications in computational 
physics.  This is reflected in the name of the 
computer, which is an acronym for {\it Computational Physics by 
Parallel Array Computer System}.  The CP-PACS has been developed in 
collaboration with Hitachi Ltd.

The CP-PACS started to operate for physics computations in April 1996 with 
1024 processing nodes.  The upgrade to the final 2048 processor system 
with a peak speed of 614 GFLOPS was completed in late September 1996.  
So far most of the CPU time has been devoted to simulations of lattice QCD.
In this article we report the performance of CP-PACS for this problem
based on the measurements recorded in the actual production runs.

We first performed a large scale simulation of QCD in the ``quenched''
approximation where the effects of quark-antiquark pair 
creation/annihilation are neglected in the intermediate processes.
Quenched QCD calculations require a large memory size and were 
performed using the entire system of 2048 nodes.  
Physics results of the quenched simulation have been presented 
elsewhere\cite{to_be_qQCD}. 
We then started a systematic study of ``full QCD'', 
progressively eliminating the quenched approximation. 
Full QCD simulations demand much more computer time than quenched 
simulations. 
Preliminary physics results of our full QCD simulations have been 
presented in \cite{compara,to_be_fQCD}. 
For a short summary of physics results from the CP-PACS, 
see \cite{Ukawa99}.

Summarizing the results for the performance of the entire CP-PACS system, 
our optimized code has achieved a sustained 
speed of 237.5 G\Fl\ for the heat-bath update of gluon variables, 
264.6 G\Fl\ for the over-relaxation update, and 325.3 G\Fl\ for 
quark matrix inversion with the even-odd preconditioned minimal residual 
algorithm.


\section{CP-PACS computer}

\begin{figure}
\begin{center} \leavevmode
\epsfxsize=12.5cm \epsfbox{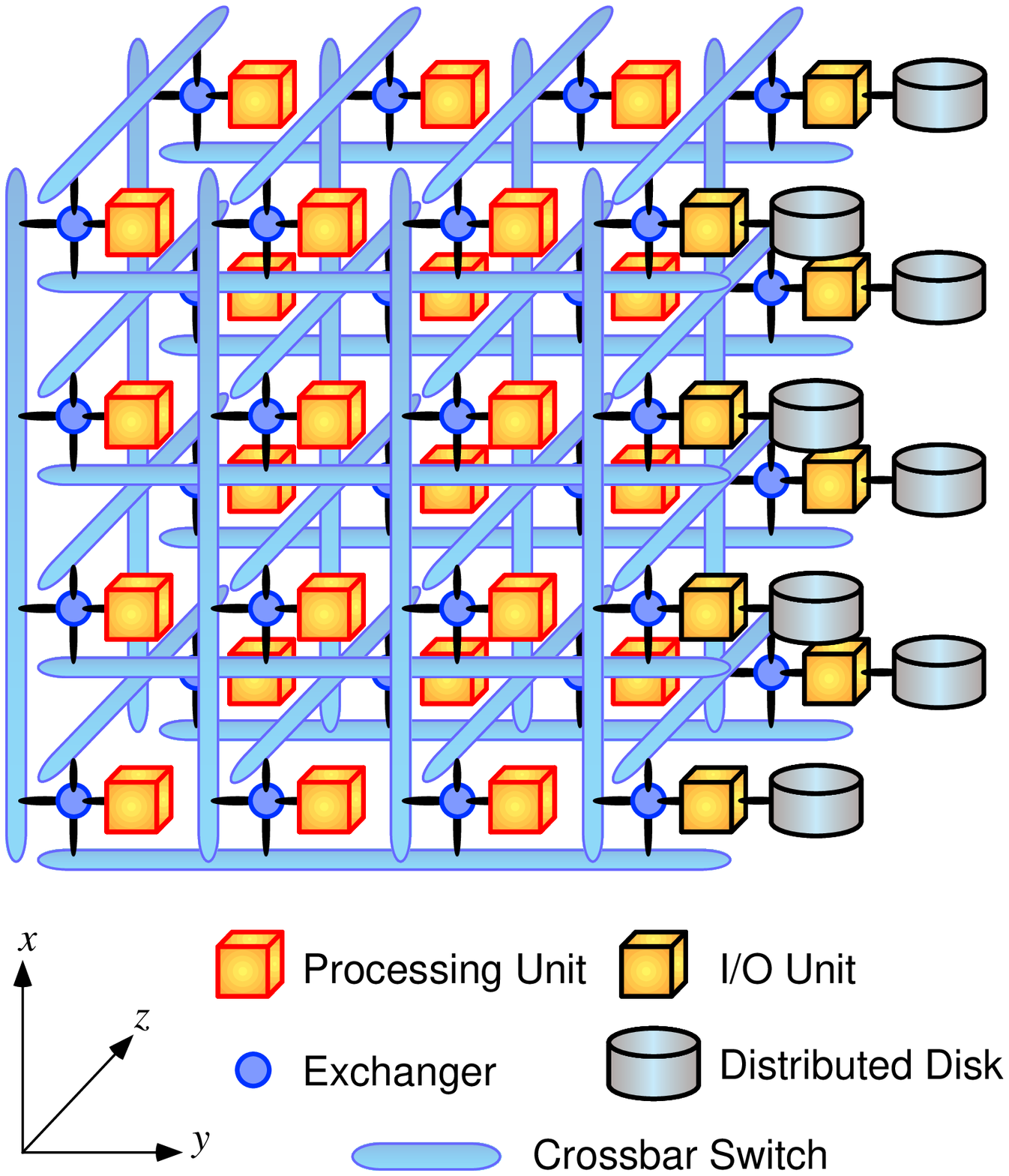}
\end{center}
\caption{A schematic diagram of the CP-PACS computer.}
\label{fig:HXB}
\end{figure}

The CP-PACS is a MIMD parallel computer with distributed memory 
consisting of 2048 processing nodes (PU) and 128 I/O nodes (IOU). 
The nodes are interconnected into an $8\times 17 \times 16$ array 
by a 3-dimensional {\it hyper-crossbar} network of 
crossbar switches as shown schematically in Figure~\ref{fig:HXB}.
Each PU has a newly made RISC processor {\it HARP1-E} 
with the peak speed of 300 MFLOPS for 64 bit data
and 64--256 MByte of main memory. 
For intermediate storage a RAID-5 disk of 8.3 GByte is attached to 
each IOU.  
Thus 
CP-PACS as a whole has the peak speed of 614 G\Fl, 
320 GByte of main memory and 1060 GByte of distributed disk space.  
We list further specifications in Table~\ref{specification}.

\begin{table}[bt]
\caption{Specification of the CP-PACS computer}
\label{specification}
\vspace{1mm}
\begin{center}
\begin{tabular}{l l}
\hline
peak speed &614 G\Fl\ (64 bit data)\\
main memory & 320 GB \\
parallel architecture &MIMD with distributed memory\\
number of nodes&2048 PU + 128 IOU\\
node processor&HARP1-E\\
\hspace*{5mm}peak speed&300 M\Fl\ (64 bit data)\\
\hspace*{5mm}architecture&HP PA-RISC1.1 + PVP-SW\\
\hspace*{5mm}\#FP registers&128\\
\hspace*{5mm}clock cycle & 150MHz\\
\hspace*{5mm}1st level cache&16KB(I) + 16KB(D)\\
\hspace*{5mm}2nd level cache&512KB(I) + 512KB(D)\\
network&3-d hyper-crossbar\\
\hspace*{5mm}node array &$8\times 17^{*}\times 16$\\
\hspace*{5mm}throughput&300MB/sec\\
\hspace*{5mm}latency    & $\leq 3 \mu$sec (hardware+software)\\
distributed disks&3.5'' RAID-5 disk (total 1060GB)\\
software&\\
\hspace*{5mm}OS&UNIX micro kernel\\
\hspace*{5mm}language & FORTRAN, C, C++, assembler\\
\hspace*{5mm} & PVM, MPI, Parallelware\\
front end& main frame connected by HIPPI\\
\hline
    & \hspace{3cm} ${}^{*}$ including IOU \\
\end{tabular}
\end{center}
\end{table}

It is well known that ordinary microprocessors cannot achieve high 
performance in large
scientific/engineering applications because cache memory does not work
effectively.  In order to solve the problem of long memory access latency 
which becomes manifest in these applications, the processor of 
CP-PACS is enhanced by a feature called {\it PVP-SW (pseudo-vector processing 
based on slide windowed registers)} \cite{PVPSW}.  This feature 
consists of: (i) a large number of floating point registers (128 in 
the actual implementation), (ii) two new instructions which request 
data preloading/poststoring to/from these registers, 
and (iii) pipelined access to main memory.  
Long memory access latency can be hidden by issuing data 
preload instructions well in advance of data usage.  Multiple data 
preloading operations are executed in a pipelined way.  The floating 
point registers are addressed with the {\it slide window} mechanism
so that the enhancement has upward compatibility with the 
PA-RISC 1.1 architecture 
which is taken as the base architecture of the CP-PACS processor.

The hyper-crossbar network has the advantage that it allows 
data transfer from a node to any other node in at most three steps, 
and that the system can be divided into independent subsystems easily. 
Internode communication is enhanced by a feature named 
{\it remote DMA} that executes direct transfer of data between user memory 
spaces of communicating nodes without using time-consuming system calls.  
This feature supports the transfer of block-strided data 
(blocks of continuous data separated by a constant stride)
as well as continuous data.

The operating system of CP-PACS is UNIX.  Each PU has a micro kernel based 
on Mach 3.0 and the IOU's have UNIX servers.  The programming languages are
Fortran 90, C, C++ and assembler.  

The loop-code optimization technique of the compiler is based on a
software pipelining technique\cite{PVPSW}.
Since a sliding of the window changes all the register numbers simultaneously, 
register allocation can be resolved flexibly.
This property makes the optimization technique simple and effective.

We refer to \cite{boku} for further details of the architecture of 
the CP-PACS and its basic performances.


\section{Coding lattice QCD on CP-PACS} 

The basic dynamical variables of QCD are the gluon field and the quark field. 
On a 4-dimensional lattice of a size $N_x \times N_y \times N_z \times N_t$, 
the gluon field is represented by a set of complex $3\times3$ matrices 
$U(n,\mu)$ where $n=(n_x,n_y,n_z,n_t)$ denotes lattice sites 
with $1\leq n_{x,y,z,t}\leq N_{x,y,z,t}$ and $\mu=x, y, z, t$ 
the four directions.  The quark field in the Wilson's formalism is 
represented by a 12-component complex vector $Q(n)$.  
The objective of lattice QCD simulations is to numerically evaluate by 
a Monte Carlo method the Feynman path integral 
\be
\langle {\cal O} \rangle
= \frac{\int {\cal O}(\bar Q, Q, U) e^{-S(\bar Q, Q, U)} 
\prod_n d\bar Q(n)dQ(n)\prod_{n,\mu} d U(n,\mu)}
{\int e^{-S(\bar Q, Q, U)} \prod_n d\bar Q(n)dQ(n)\prod_{n,\mu} d U(n,\mu)}
\ee
where ${\cal O}(\bar Q, Q, U)$ is a physical observable 
and $S(\bar Q, Q, U)$ the action of lattice QCD describing the interaction 
of quarks and gluons.
Typically the main simulation steps consist of the following:
\begin{itemize} 
\begin{enumerate} 
\item {\bf Update} of the gluon configuration $\{U(n,\mu)\}$ to generate 
the distribution $e^{-S}$. 
\item {\bf Gauge fixing} to reduce statistical noise in the measurement of 
      observables. 
\item {\bf Solver} to compute the quark propagator 
$
G = D^{-1}(U) \, B
$
for a number of given quark sources $B$, 
where the quark matrix $D(U)$ is a sparse $12V \times 12V$ complex matrix 
depending on the gluon configuration and 
$V=N_x\cdot N_y\cdot N_z\cdot N_t$. 
\item {\bf Measurement} of hadron observables by combining quark propagators.
\end{enumerate}
\end{itemize}
The whole cycle is repeated 
several hundred to several thousand times.
The algorithm for the update part differs significantly between  
{\it quenched} simulations, 
in which the back reaction of quark fields on the distribution of gluon 
fields is ignored, and {\it full QCD} simulations without such an 
approximation.  
For full QCD, the most time-consuming computation in the update part is 
the same as that in the solver part: $G = D^{-1}(U) \, B$.
For both cases the computer time is mostly spent in the update 
part and the solver part, with the update part weighted dominantly for 
full QCD simulations.

Our performance evaluation data come from a quenched simulation
using the entire 2048 PU's of the CP-PACS, and from a full QCD
simulation using a sub-partition consisting of 512 PU's.

For the quenched simulation, 
the update part is carried out with the heat-bath \cite{HB} and 
the over-relaxation \cite{OR} algorithms 
mixed in a ratio of 1:4. 
For the full QCD simulation we adopt the hybrid Monte Carlo 
algorithm \cite{HMC}.

The computation of $G = D^{-1}(U) \, B$ is performed with 
a minimal residual algorithm or a BiCGStab algorithm \cite{BiCGStab}.
In the both cases, even-odd preconditioning can be applied.
For the quenched simulation we adopt the former, while for full QCD
mostly the latter.

Our QCD codes for the CP-PACS are originally developed in 
FORTRAN90.  
Let $\Ndx \times \Ndy \times \Ndz$ be the 3-dimensional PU
array.  We divide the total lattice
into $\Ndx \times \Ndy \times \Ndz$ sublattices, each with a size 
$L_x \times L_y \times L_z \times N_t$ with $L_x=N_x/\Ndx$  {\it etc.}
On each PU the gluon field is then defined as  
\be
{\tt complex\!*\!16} \,\,\,
{\tt U(3,3,0:}N_t{\tt +1,0:}L_z{\tt +1,0:}L_y{\tt +1,0:}L_x{\tt +1,4)},
\ee
for the case of the standard one plaquette gauge action, and
\be
{\tt complex\!*\!16} \,\,\,
{\tt U(3,3,-1:}N_t{\tt +2,-1:}L_z{\tt +2,-1:}L_y{\tt +2,-1:}L_x{\tt +2,4)},
\ee
for the case of the improved gauge action 
\cite{Iwasaki}
we adopted in full QCD simulations.
For the quark field (or quark propagator to be precise) we introduce
\be
{\tt complex\!*\!16} \,\,\,
{\tt G(3,4,0:}N_t{\tt +1,0:}L_z{\tt +1,0:}L_y{\tt +1,0:}L_x{\tt +1)}.
\ee
These arrays contain boundaries 
({\tt ix} $\leq 0$ and $\geq L_x+1$, {\it etc.}),
which are the copies of the corresponding variables on the neighboring PU's.
After each modification of {\tt U} or {\tt G}, the boundary values have to be 
renewed.  The library functions for a block-strided remote DMA transfer 
enable us to perform the necessary boundary copies efficiently without 
gather/scatter manipulations. 

A characteristic feature of computations in QCD 
is that the number of load/store operations, multiplications, and additions 
are approximately the same.  For the heat-bath and over-relaxation updates
in quenched QCD, for example, the dominant computation is a matrix 
multiplication of two gluon variables of form {\tt U*U} with the two 
{\tt U}'s on neighboring links. For each column of the resulting matrix,
this computation requires 24 loads, 6 stores, 36 multiplications and 30 additions.  
The superscalar feature of the CP-PACS processor includes a simultaneous
multiplication-addition instruction which can be issued concurrently with 
a preload/poststore operation. A poststore operation requires two machine cycles
on the CP-PACS. Thus computations of the above type can be 
effectively carried out.  
The PVP-SW feature solves the remaining problem of 
large memory access latency and the necessity of a large number of registers 
to handle a large number of data per loop index.  
Our compiler can 
schedule instructions for these loops well.  With an appropriate choice 
of loop unrolling, the FORTRAN code achieves over 160 M\Fl/PU
for the loop, which is over 50\% of peak speed.

\begin{figure}
\begin{center} \leavevmode
\epsfxsize=12cm \epsfbox{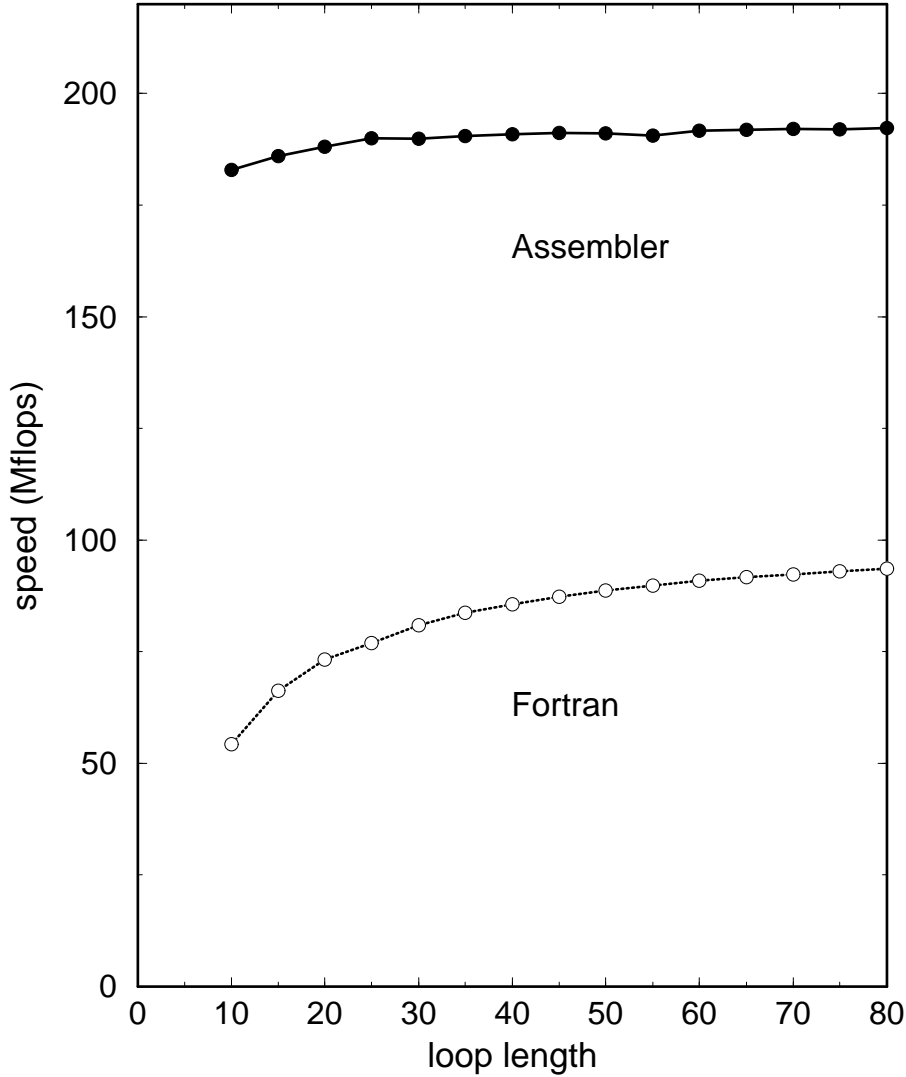}
\end{center}
\caption{Performance/PU of the core subroutine {\tt mult} of the solver part 
with assembler coding 
as a function of loop length.
In our QCD code, the loop length is $N_t/2$ due to an even-odd algorithm. 
The peak speed of a PU is 300 M\Fl 
when the numbers of multiplications, additions, and loads/stores are equal. 
Because of an imbalance of these numbers,
the theoretical peak speed of this loop is 266 M\Fl. 
For comparison, we also plot performance of our fastest FORTRAN 
code for the same calculation.
}
\label{fig:mult}
\end{figure}

In the the computation of $G = D^{-1}(U) \, B$ with an iterative algorithm 
such as minimal residual and BiCGStab,
the dominant computation is executed in a subroutine called {\tt mult} which 
computes {\tt G}$\to${\tt D(U)*G}.
A main part of the calculation, which may be schematically written as
{\tt V}={\tt U*(G$_1$+G$_2$)}, is 
well balanced requiring 30 loads, 6 stores, 36 multiplications and 36 
additions.  The remaining computation is less balanced.
An efficient scheduling of all these instructions is a fairly complicated 
task, which the present compiler can cope with only to a modest 
degree (see Fig.~\ref{fig:mult}).  
Since an efficient solver is a very important ingredient of lattice 
QCD simulations, we have written this part of the code 
in the assembly language \cite{yoshieY}. 
All the instructions are hand-optimized 
within two long loops paying attention to remove unnecessary load/store 
operations using the PVP-SW feature. 
The performance of the assembler code exceeds 
190 M\Fl/PU as shown in Fig.~\ref{fig:mult}. 

We emphasize that the loop length for half the peak performance is 
quite short for our processor, especially for the assembler code.  This 
feature, originating from the pseudo-vectorization on a RISC processor 
by the software pipelining technique, 
is in a marked contrast with vector computers; they generally 
require a long loop length for peak performance, which is usually realized 
by a one-dimensional addressing of the multi-dimensional coordinate $n$.  
In our case, we use only the $t$-coordinate for pseudo-vectorization. 
This made the program quite easy to read and debug.

\begin{table}[bt]
\caption{Execution time of a single cycle of a quenched QCD simulation 
to study hadron spectroscopy.  The lattice size is $64^3\times112$. 
For the update part, we performed 4 OR sweeps every one HB update. 
Quark propagators for five different quark masses are solved 
at the solver part, and the results stored on distributed disks are 
collected in the measurement part to combine hadron observables.
}
\label{tab:fraction}
\vspace{1mm}
\begin{center}
\begin{tabular}{lcc}
\hline
  & execution time (hour) & fraction (\%)\\
\hline
update              & 4.8 & 30 \\
\hspace{5mm}heat-bath       & (1.6) & (10) \\
\hspace{5mm}over-relaxation & (3.0) & (19) \\
gauge-fixing        & 1.3 &  8 \\
solver              & 6.8 & 43 \\
measurement         & 2.9 & 19 \\
\hline
total               & 15.7 & 100 \\
\hline
\end{tabular}
\end{center}
\end{table}

\section{Performance of quenched simulation}

We now describe the performance of our QCD programs on the CP-PACS 
computer obtained in our production run for a quenched QCD hadron spectrum 
calculation with the full 2048 processing units.  The lattice size is  
$64^3\times 112$ which is the largest lattice employed so far in lattice QCD 
simulations.
In Table~\ref{tab:fraction} we list a breakdown of the execution time of 
one cycle of simulation into those of the four basic steps.
During the run we measure the execution times of important subroutines 
using a FORTRAN-supplied interval timer {\tt xclock} which has the 
precision of micro seconds.  
Performance data reported below is based on this measurement averaged over 
20 cycles.

\begin{table}[tb]
\caption{Profile of the solver program running on the CP-PACS 
with 2048 PU. 
The subroutines {\tt rbmr} and {\tt mult} are coded in
assembler.
{\tt rbmr} is the main routine excluding the subroutine {\tt mult}, and 
{\tt copyG} and {\tt gsum} 
executes data transfer and global sum, respectively.
The number of floating point operations in {\tt gsum} is negligible and
is ignored.  
}
\label{tab:solverCPPACS}
\vspace{1mm}
\begin{center}
\begin{tabular}{ccccc}
\hline
subroutine & time/call (sec) & \#flop & M\Fl/PU &
ratio of exec.time \\
\hline
{\tt rbmr}  & 0.0163 &  3096576 & 189.8 &  0.120 \\
{\tt mult}  & 0.0482 &  9289728 & 192.5 &  0.707 \\
{\tt copyG} & 0.0098 &  0       & 0     &  0.144 \\
{\tt gsum}  & 0.0020 &  0       & 0     &  0.029 \\
\hline
solver in total &      &          & 158.8 & \\
\hline
\end{tabular}
\end{center}
\end{table}

\subsection{Update of gluon configuration}

In quenched QCD simulations, the performance of a gluon update program is 
often compared 
in terms of the ``link update time'', {\it i.e.,} the execution time needed
to update a single gluon field $U(n,\mu)$ on a link $(n,\mu)$.

The link update time may be translated into GFLOPS using the number of 
floating operations per link.  For the heat bath update, this number 
cannot be fixed uniquely since the heat bath is a stochastic 
process and elementary functions are called in the program.  
We adopt the number 5700 which is widely accepted in the lattice community 
and has been used in previous estimations\cite{NWT}. 

The previous best performance among published data was obtained 
by the NAL-Yamagata collaboration\cite{NWT}.  Employing the Numerical 
Wind Tunnel (NWT) with 128 nodes at National Aerospace Laboratory (NAL), 
they reported a link update time of 0.0317 $\mu$sec or 179.8 G\Fl, 
which won the Gordon Bell Prize of 1995. 
A better performance is listed on the web page of NAL\cite{NWT2}:
215.8 G\Fl\ obtained on the enhanced NWT with 160 nodes, 
corresponding to the link-update time 0.0264 $\mu$sec.
Performance on other computers is summarized in \cite{NWT}. 

The link update time on the CP-PACS is 0.0240 $\mu$sec.
This is equivalent to 237.5 G\Fl, which exceeds the best value 
on NWT by 10\%.

For the over-relaxation update, our link-update time is 0.0112 $\mu$sec.
The number of floating point operations per link can be counted precisely
for this case, which equals 3050 using the algorithm we adopted \cite{OR}. 
Therefore, our over-relaxation code achieves 264.6 G\Fl. 

\begin{table}[tb]
\caption{Solver performance in the literature}
\label{tab:solverOther}
\vspace{1mm}
\begin{center}
\begin{tabular}{c|cccccc}
\hline\hline
\multicolumn{7}{l}{Reported performance data} \\ 
\hline
  Machine      &  VPP500 &  CM-5   &  Paragon &  T3D  &  CM-5 &  ACPMAPS   \\
               & (Fujitsu) & (TMC) & (Intel) & (Cray) & (TMC) & \\
\hline                                    
  location     &  KEK$^a$ & LANL$^b$ & --- & PSC$^c$ & --- &  Fermilab$^d$  \\
  \#node       &  64     &  ---    &  64      &  64   &  64   &  128       \\
  Problem      &  KS     &  Wilson &  KS      &  KS   &  KS   &  Wilson    \\
  M\Fl/node  &  1105   &  35     &  23.2    &  22.2 &  20.0 &  8.5       \\
  G\Fl       &  70.7   &  ---    &  1.5     &  1.4  &  1.3  &  1.1       \\
  comment      &  (1)    &  (2)    &  (3)     &  (3)  &  (3)  &  (3)       \\
  reference    &\cite{KEK}&\cite{LANL}&  \cite{MILC}  &  \cite{MILC} &  \cite{MILC} & 
\cite{Fermilab}   \\
\hline\hline
\multicolumn{7}{l}{Possible performance} \\ 
\hline
  Machine      &  NWT    &  CM-5   &  Paragon & T3D    & CM-5 & ACPMAPS    \\
\hline                                      
  location     & NAL$^e$ & LANL$^b$ & SNL$^f$ & US Gov.& LANL$^b$ & Fermilab$^d$   \\
  \#node       &  167    &  1056   &  3680    & 1024   & 1056 & 612        \\
  G\Fl       &  196.1  &  37.0   &  85.4    & 22.7   & 21.1 & 5.2        \\
\hline\hline
\multicolumn{7}{l}{(a) KEK: National Laboratory for High Energy Physics, Japan.} \\
\multicolumn{7}{l}{(b) LANL: Los Alamos National Laboratory, USA} \\
\multicolumn{7}{l}{(c) PSC: Pittsburgh Supercomputing Center, USA} \\
\multicolumn{7}{l}{(d) Fermilab: Fermi National Accelerator Laboratory, USA} \\
\multicolumn{7}{l}{(e) NAL: National Aerospace Laboratory, Japan}\\
\multicolumn{7}{l}{(f) SNL: Sandia National Laboratories, USA}\\
\multicolumn{7}{l}{(1) performance of the core part (mult) of the solver} \\
\multicolumn{7}{l}{(2) performance including I/O and setup overhead, 
MR4 algorithm} \\
\multicolumn{7}{l}{(3) conjugate gradient algorithm} \\
\end{tabular}
\end{center}
\end{table}

\subsection{Quark propagator solver} \label{sec:solver}

The calculation of performance for the quark propagator solver is
made as follows.  
The computations in the main loop consist of four subroutines 
(see Table~\ref{tab:solverCPPACS}).
For each subroutine, we measure 
the execution time
per one subroutine call and the ratio of time spent for the subroutine 
in the total execution time. 
Counting the number of floating point operations 
in each subroutine, which can be done precisely, we convert the time data
into G\Fl. The raw data for each subroutine are summarized 
in Table~\ref{tab:solverCPPACS}.
The average of the performances with weight of the time ratio
leads to 158.8 M\Fl/PU for the solver as a whole,
or equivalently 325.3 G\Fl\ for the entire CP-PACS system. 

For comparison, 
we reproduce in the top half of Table~\ref{tab:solverOther} 
the performance data for several machines reported in the literature.  
They refer to either the Wilson quark action used in our run or the 
Kogut-Susskind quark action which is another form of lattice quark action
often used in lattice QCD simulations.  The algorithm
for the solver also differs as noted in the comments.
We should remark in particular that these results are mostly calculated by
employing a smaller number of nodes than maximally available.
Therefore, we have tried to estimate ``possible performance'' by 
assuming that the measured performance scales linearly with the 
number of nodes up to the maximal configuration with the same 
architecture which exists now or has existed in the past.  
This estimate is shown on the bottom half of 
Table~\ref{tab:solverOther}.

The best performance for a quark matrix solver recorded has been obtained 
on the VPP500 with 64 nodes for a KS quark propagator solver. 
If the reported value is translated to the possible performance
on the NWT with 167 nodes, we obtain 196.1 G\Fl, 
although NWT has not been used for hadron spectroscopy calculations.
The measured performance of 325.3 G\Fl\ on the CP-PACS is 66\% 
larger than this.

\section{Full QCD simulations}

In full QCD simulations, the most time-consuming part is {\tt mult}
both in update and solver.
The performance of {\tt mult} has already been discussed in 
Sec.~\ref{sec:solver}.
Here, however, several additional remarks are in order.

First, full QCD simulations are extremely computer-time 
consuming compared to those of quenched QCD. 
Simple scaling estimates place a hundred-fold 
or more increase in the amount of computations for full QCD 
compared to that of quenched QCD with current algorithms. 
Therefore, the use of a large lattice comparable to the quenched
case is difficult; we are forced to employ coarse lattices.

In order to keep a reasonable sublattice size for each PU, 
we perform calculations on partitions of the CP-PACS. 
The largest partition we have used in our full QCD simulation 
consists of 512 PU's with a lattice size of $24^3\times48$. 
We report the performance measured on this lattice. 
In this case, the vector length is 24 for major loops.

Furthermore, in order to suppress discretization errors caused 
by the coarse lattice used in full QCD calculations,
we apply action improvement which has been widely pursued 
in the last few years. 
Based on a comparative study of various combinations of 
improved and unimproved actions in full QCD \cite{compara},
we adopt an improved gluon action proposed by Iwasaki \cite{Iwasaki},
combined with an improved quark action suggested by Sheikholeslami and 
Wohlert \cite{clover}. 
Although the basic structure of the computations can be maintained, 
improvement of the action implies several additional computations
which are coded in FORTRAN in our programs.
Together with the short vector length, the relative weight of the
FORTRAN parts over the parts coded in the assembly language 
(such as {\tt mult}) is much larger than that in the quenched simulations.

In the hybrid Monte Carlo algorithm \cite{HMC} adopted in our full QCD 
update, one unit of update calculation is called ``trajectory''.
Several trajectories are required to suppress autocorrelation among
succeeding configurations. 
In our simulations, we separated the measurement steps by five trajectories. 

Major parts of the computer time for one trajectory of full QCD 
update can be assembled as 
\be
\left[
A_{\rm comp} + A_{\rm commun} + (B_{\rm comp} + B_{\rm commun}) 
\times N_{\rm inv} \right] \times {1 \over \Delta\tau}
+ C,
\ee
with negligible $C$.
Here, $N_{\rm inv}$ is the number of iteration steps required in the 
computation of $G = D^{-1}(U) \, B$, and 
$\Delta\tau$ is the evolution step size for the 
hybrid Monte Carlo algorithm.
In our simulations, $N_{\rm inv} \approx 60$--500
and $1/\Delta\tau \approx 50$--150
depending on the values of physical parameters 
such as quark mass and lattice volume.

$A_{\rm comp}$ and $B_{\rm comp}$ are contributions from the computations 
in each PU, whose most time-consuming part is {\tt mult}.
For the case of the algorithm we adopt, we find 
that the total number of floating point operations 
for $A_{\rm comp}$ and $B_{\rm comp}$ are
$456,000 \times V$ and $8,800 \times V$, respectively, where 
$V = N_x \times N_y \times N_z \times N_t$ is the lattice volume.
Relatively large number of operations for $A_{\rm comp}$ and $B_{\rm comp}$ 
is due to the improved action and the improved algorithm we use.
The speed of $A_{\rm comp}$ and $B_{\rm comp}$ are measured to be 
94 and 113 M\Fl/PU, respectively,
which correspond to 31 and 38\% of the peak performance.

$A_{\rm commun}$ and $B_{\rm commun}$ are from internode communications.
In our simulations, the computer time for $A_{\rm commun}$ is negligible 
and that for $B_{\rm commun}$ is about 8--10\% of $B_{\rm comp}$.
Measured throughput for the $B_{\rm commun}$ part was 157 MB/sec 
on a $24^3\times48$ 
lattice simulated with a 512 PU partition. 
An extrapolation to the limit of large $N_t$ leads to 192 MB/sec.

\section{Summary}

We have presented the performance data of the CP-PACS computer 
measured during recent production runs for Quantum Chromodynamics
simulations of quarks and gluons.
In a run with the quenched approximation, we used 
the full 2048 processing nodes of the CP-PACS and 
obtained a sustained speed of 237.5 G\Fl\ for the heat-bath update of gluon
variables, 264.6 G\Fl\ for the over-relaxation update, and 325.3 G\Fl\ for 
quark matrix inversion with an even-odd preconditioned minimal residual 
algorithm.
These performances correspond to 43--53\% of the theoretical peak
speed of the CP-PACS. 
In more recent full QCD simulations in which the quenched approximation
is removed, we used sub-partitions of the CP-PACS up to 512 processing
nodes. 
We found 113 M\Fl/PU for the kernel part of the simulation,
which corresponds to 38\% of the peak performance.

\section*{Acknowledgements}
We are grateful to members of the CP-PACS project for 
discussions and encouragements.
We thank Y.\ Iwasaki, A.\ Ukawa and H.P.\ Shanahan 
for reading the manuscript. 
This work is supported by 
the Grant-in-Aid of Ministry of Education,
Science and Culture (No.\ 08NP0101 and No.\ 10640248).


\end{document}